

%
\catcode`\@=11 

\newcount\yearltd\yearltd=\year\advance\yearltd by -1900

%
%

\def\draftmode{\message{ DRAFTMODE }\def\draftdate{{\rm preliminary draft:
\number\month/\number\day/\number\yearltd\ \ \hourmin}}%
\headline={\hfil\draftdate}\writelabels\baselineskip=20pt plus 2pt minus 2pt
 {\count255=\time\divide\count255 by 60 \xdef\hourmin{\number\count255}
  \multiply\count255 by-60\advance\count255 by\time
  \xdef\hourmin{\hourmin:\ifnum\count255<10 0\fi\the\count255}}}
\def\nolabels{\def\wrlabel##1{}\def\eqlabel##1{}\def\reflabel##1{}}
\def\writelabels{\def\wrlabel##1{\leavevmode\vadjust{\rlap{\smash%
{\line{{\escapechar=` \hfill\rlap{\sevenrm\hskip.03in\string##1}}}}}}}%
\def\eqlabel##1{{\escapechar-1\rlap{\sevenrm\hskip.05in\string##1}}}%
\def\thlabel##1{{\escapechar-1\rlap{\sevenrm\hskip.05in\string##1}}}%
\def\reflabel##1{\noexpand\llap{\noexpand\sevenrm\string\string\string##1}}}
\nolabels
%
\global\newcount\secno \global\secno=0
\global\newcount\meqno \global\meqno=1
\global\newcount\mthno \global\mthno=1
\global\newcount\mexno \global\mexno=1

\def\newsec#1{\global\advance\secno by1\message{(\the\secno. #1)}
\global\subsecno=0\xdef\secsym{\the\secno.}\global\meqno=1\global\mthno=1
\global\mexno=1
\bigbreak\bigskip\noindent{\bf\the\secno. #1}\writetoca{{\secsym} {#1}}
\par\nobreak\medskip\nobreak}
\xdef\secsym{}
\global\newcount\subsecno \global\subsecno=0
\def\subsec#1{\global\advance\subsecno by1\message{(\secsym\the\subsecno. #1)}
\bigbreak\noindent{\it\secsym\the\subsecno. #1}\writetoca{\string\quad
{\secsym\the\subsecno.} {#1}}\par\nobreak\medskip\nobreak}
\def\appendix#1#2{\global\meqno=1\global\mthno=1\global\mexno=1
\global\subsecno=0
\xdef\secsym{\hbox{#1.}}
\bigbreak\bigskip\noindent{\bf Appendix #1. #2}\message{(#1. #2)}
\writetoca{Appendix {#1.} {#2}}\par\nobreak\medskip\nobreak}
%
%
\def\eqnn#1{\xdef #1{(\secsym\the\meqno)}\writedef{#1\leftbracket#1}%
\global\advance\meqno by1\wrlabel#1}
\def\eqna#1{\xdef #1##1{\hbox{$(\secsym\the\meqno##1)$}}
\writedef{#1\numbersign1\leftbracket#1{\numbersign1}}%
\global\advance\meqno by1\wrlabel{#1$\{\}$}}
\def\eqn#1#2{\xdef #1{(\secsym\the\meqno)}\writedef{#1\leftbracket#1}%
\global\advance\meqno by1$$#2\eqno#1\eqlabel#1$$}
%
%
\def\thm#1{\xdef #1{\secsym\the\mthno}\writedef{#1\leftbracket#1}%
\global\advance\mthno by1\wrlabel#1}
\def\exm#1{\xdef #1{\secsym\the\mexno}\writedef{#1\leftbracket#1}%
\global\advance\mexno by1\wrlabel#1}
%
\newskip\footskip\footskip14pt plus 1pt minus 1pt 
\def\f@@t{\baselineskip\footskip\bgroup\aftergroup\@foot\let\next}
\setbox\strutbox=\hbox{\vrule height9.5pt depth4.5pt width0pt}
\global\newcount\ftno \global\ftno=0
\def\foot{\global\advance\ftno by1\footnote{$^{\the\ftno}$}}
%
\newwrite\ftfile
\def\footend{\def\foot{\global\advance\ftno by1\chardef\wfile=\ftfile
$^{\the\ftno}$\ifnum\ftno=1\immediate\openout\ftfile=foots.tmp\fi%
\immediate\write\ftfile{\noexpand\smallskip%
\noexpand\item{f\the\ftno:\ }\pctsign}\findarg}%
\def\footatend{\vfill\eject\immediate\closeout\ftfile{\parindent=20pt
\centerline{\bf Footnotes}\nobreak\bigskip\input foots.tmp }}}
\def\footatend{}
%
%
\global\newcount\refno \global\refno=1
\newwrite\rfile
\def\ref{\the\refno\nref}
\def\bref{\nref}
\def\nref#1{\xdef#1{\the\refno}\writedef{#1\leftbracket#1}%
\ifnum\refno=1\immediate\openout\rfile=refs.tmp\fi
\global\advance\refno by1\chardef\wfile=\rfile\immediate
\write\rfile{\noexpand\item{[#1]\ }\reflabel{#1\hskip.31in}\pctsign}\findarg}
\def\findarg#1#{\begingroup\obeylines\newlinechar=`\^^M\pass@rg}
{\obeylines\gdef\pass@rg#1{\writ@line\relax #1^^M\hbox{}^^M}%
\gdef\writ@line#1^^M{\expandafter\toks0\expandafter{\striprel@x #1}%
\edef\next{\the\toks0}\ifx\next\em@rk\let\next=\endgroup\else\ifx\next\empty%
\else\immediate\write\wfile{\the\toks0}\fi\let\next=\writ@line\fi\next\relax}}
\def\striprel@x#1{} \def\em@rk{\hbox{}}

\def\addref#1{\immediate\write\rfile{\noexpand\item{}#1}} 
\def\startrefs#1{\immediate\openout\rfile=refs.tmp\refno=#1}
\def\xref{\expandafter\xr@f}\def\xr@f[#1]{#1}
\def\refs#1{[\r@fs #1{\hbox{}}]}
\def\r@fs#1{\edef\next{#1}\ifx\next\em@rk\def\next{}\else
\ifx\next#1\xref #1\else#1\fi\let\next=\r@fs\fi\next}
%

%
\newwrite\ffile\global\newcount\figno \global\figno=1
\def\fig{fig.~\the\figno\nfig}
\def\nfig#1{\xdef#1{fig.~\the\figno}%
\writedef{#1\leftbracket fig.\noexpand~\the\figno}%
\ifnum\figno=1\immediate\openout\ffile=figs.tmp\fi\chardef\wfile=\ffile%
\immediate\write\ffile{\noexpand\medskip\noexpand\item{Fig.\ \the\figno. }
\reflabel{#1\hskip.55in}\pctsign}\global\advance\figno by1\findarg}
\def\xfig{\expandafter\xf@g}\def\xf@g fig.\penalty\@M\ {}
\def\figs#1{figs.~\f@gs #1{\hbox{}}}
\def\f@gs#1{\edef\next{#1}\ifx\next\em@rk\def\next{}\else
\ifx\next#1\xfig #1\else#1\fi\let\next=\f@gs\fi\next}
\newwrite\lfile
{\escapechar-1\xdef\pctsign{\string\%}\xdef\leftbracket{\string\{}
\xdef\rightbracket{\string\}}\xdef\numbersign{\string\#}}

\def\writestop{\def\writestoppt{\immediate\write\lfile{\string\pageno%
\the\pageno\string\startrefs\leftbracket\the\refno\rightbracket%
\string\def\string\secsym\leftbracket\secsym\rightbracket%
\string\secno\the\secno\string\meqno\the\meqno}\immediate\closeout\lfile}}
\def\writestoppt{}\def\writedef#1{}
\def\seclab#1{\xdef #1{\the\secno}\writedef{#1\leftbracket#1}\wrlabel{#1=#1}}
\def\subseclab#1{\xdef #1{\secsym\the\subsecno}%
\writedef{#1\leftbracket#1}\wrlabel{#1=#1}}
\newwrite\tfile \def\writetoca#1{}
\def\leaderfill{\leaders\hbox to 1em{\hss.\hss}\hfill}
\def\writetoc{\immediate\openout\tfile=toc.tmp
   \def\writetoca##1{{\edef\next{\write\tfile{\noindent ##1
   \string\leaderfill {\noexpand\number\pageno} \par}}\next}}}

%
\catcode`\@=12 
%
\ifx\answ\bigans
 
 \font\titlei=cmmi10 scaled\magstep3
\font\titleis=cmmi7 scaled\magstep3 \font\titleiss=cmmi5 scaled\magstep3
\font\titlesy=cmsy10 scaled\magstep3 \font\titlesys=cmsy7 scaled\magstep3
\font\titlesyss=cmsy5 scaled\magstep3 
\else
 
 \font\titlei=cmmi10 scaled\magstep4
\font\titleis=cmmi7 scaled\magstep4 \font\titleiss=cmmi5 scaled\magstep4
\font\titlesy=cmsy10 scaled\magstep4 \font\titlesys=cmsy7 scaled\magstep4
\font\titlesyss=cmsy5 scaled\magstep4 
 
 \font\absi=cmmi10 scaled\magstep1
\font\absis=cmmi7 scaled\magstep1 \font\absiss=cmmi5 scaled\magstep1
\font\abssy=cmsy10 scaled\magstep1 \font\abssys=cmsy7 scaled\magstep1
\font\abssyss=cmsy5 scaled\magstep1 
\skewchar\absi='177 \skewchar\absis='177 \skewchar\absiss='177
\skewchar\abssy='60 \skewchar\abssys='60 \skewchar\abssyss='60
\fi
\skewchar\titlei='177 \skewchar\titleis='177 \skewchar\titleiss='177
\skewchar\titlesy='60 \skewchar\titlesys='60 \skewchar\titlesyss='60
\ifx\answ\bigans\else
 \fi
%

%
%
%

\def\p{\partial}

\def\vev#1{\langle #1 \rangle}

\def\darr#1{\raise1.5ex\hbox{$\leftrightarrow$}\mkern-16.5mu #1}
\def\half{{\textstyle{1\over2}}} 
%
%
\def\al{\alpha}
\def\be{\beta}
  \def\Ga{\Gamma}
\def\de{\delta}

\def\ph{\phi}

  \def\Om{\Omega}
%
%

%

%
%

\def\cF{{\cal F}}

\def\cL{{\cal L}}
\def\cM{{\cal M}}

\def\cU{{\cal U}}

%
%
%
\def\lefthook{{\vrule height5pt width0.4pt depth0pt}}
\def\righthook{{\vrule height5pt width0.4pt depth0pt}}
\def\leftrighthookfill{$\mathsurround=0pt \mathord\lefthook
     \hrulefill\mathord\righthook$}
\def\underhook#1{\vtop{\ialign{##\crcr$\hfil\displaystyle{#1}\hfil$\crcr
      \noalign{\kern-1pt\nointerlineskip\vskip2pt}
      \leftrighthookfill\crcr}}}
%
%


\def\Box{\rlap{$\sqcup$}$\sqcap$}

\def\ie{{\it i.e.\ }}
\def\eg{{\it e.g.\ }}

\def\proof{\noindent {\it Proof:}\ }

\def\ZZ{Z\!\!\!Z}               
                 %
\def\CC{I\!\!\!\!C}             %

\def\bs{\bigskip}



\def\mapright#1{\smash{\mathop{\longrightarrow}\limits^{#1}}}

%

\def\con#1{[\![\,#1\,]\!]}


\magnification=1200

  \def\cM{{\cal M}}

\def\mod#1{{\rm mod}\,#1}
\def\hh#1{h_{#1}} \def\pp#1{p_{#1}}
\def\ot{\leftarrow} \def\to{\rightarrow}
\nopagenumbers\pageno=0
\line{\hfill{CERN-TH.6196/91}}
\bigskip
\centerline{{\bf FOCK SPACE RESOLUTIONS OF THE VIRASORO}}
\centerline{{\bf HIGHEST WEIGHT MODULES WITH $c\leq  1$ }}
\bigskip\bigskip
\centerline{Peter Bouwknegt}
\smallskip
\centerline{{\it CERN - Theory Division}}
\centerline{{\it CH-1211 Geneva 23}}
\centerline{{\it Switzerland}}
\bigskip
\centerline{Jim McCarthy\footnote{$^{\dagger}$}{Supported by the NSF Grant
\#PHY-88-04561.}}
\smallskip
\centerline{{\it Department of Physics}}
\centerline{{\it Brandeis University}}
\centerline{{\it Waltham, MA 02254}}
\bigskip
\centerline{{ Krzysztof Pilch\footnote{$^\ddagger$}{Supported
in part by the Department of Energy Contract \#DE-FG03-84ER-40168 and
by the USC Faculty Research and Innovation Fund.}}}
\smallskip
\centerline{{\it Department of Physics}}
\centerline{{\it University of Southern California}}
\centerline{{\it Los Angeles, CA 90089-0484}}
\bigskip
\bigskip
\centerline{{\bf Abstract}}\smallskip

We extend Felder's construction of Fock space resolutions for the
Virasoro minimal models to all irreducible modules with $c\leq 1$.
In particular, we provide resolutions for the representations corresponding
to the boundary and exterior of the Kac table.

\bs\bs
\centerline{Submitted to: {\it Letters in Mathematical Physics}}
\vfil
\line{ CERN-TH.6196/91   \hfill August 1991}
\line{ BRX TH-323 \hfill revised August 1991}
\line{ USC-91/21  \hfill}

\eject
\footline={\hss\tenrm\folio\hss}

\baselineskip=1.2\baselineskip

\def\NP{Nucl. Phys.\ }

\def\CMP{Commun. Math. Phys.\ }

\def\LMP{Lett. Math. Phys.\ }
\def\LNM{Lect. Notes Math.\ }

\bref\Fe{
G. Felder, Nucl. Phys. {\bf B317} (1989) 215;
erratum, {\it ibid.} {\bf B324} (1989) 548.}
\bref\BPZ{
A.A. Belavin, A.M. Polyakov and A.B. Zamolodchikov,
Nucl. Phys. {\bf B241} (1984) 333.}
\bref\BMPrev{
P. Bouwknegt, J. McCarthy, and K. Pilch, Prog. Theor. Phys. Suppl. {\bf 102}
(1990) 67.}
\bref\FFear{B.L. Feigin and D.B. Fuchs,
Funct. Anal. Appl. {\bf 16} (1982) 114; {\it ibid.}
{\bf 17} (1983) 241; \LNM {\bf 1060} (1984) 230.}
\bref\FF{
B.L. Feigin and D.B. Fuchs,
in {\it Representations of infinite-dimensional Lie groups and
Lie algebras}, Gordon and Breach,
New York (1989).}
\bref\Thorn{
C. Thorn, \NP {\bf B248} (1984) 551.}
\bref\FFK{
G. Felder, J. Fr\"ohlich, and G. Keller, \CMP {\bf 124} (1989) 647. }
\bref\TK{
A. Tsuchiya and Y. Kanie, Publ. RIMS {\bf 22} (1986) 259.}
\bref\FD{
Vl.S. Dotsenko and V.A. Fateev, Nucl. Phys. {\bf B251} [FS13] (1985) 691.}
\bref\Ka{
V.G. Kac, \LNM {\bf 94} (1979) 441.}
\bref\GSW{
M.B. Green, J.H. Schwarz, and E. Witten, {\it Superstring Theory}, Cambridge
University Press, Cambridge (1987).}
\bref\FMS{
D. Friedan, E. Martinec, and S. Shenker, \NP {\bf B271} (1986) 93.}
\bref\BMPcmp{
P. Bouwknegt, J. McCarthy, and K. Pilch, \CMP {\bf 131} (1990) 125.}
\bref\WY{
M. Wakimoto and H. Yamada, \LMP {\bf 7} (1983) 513.}
\bref\KR{
V.G. Kac and A.K. Raina, Adv. Ser. Math. Phys. {\bf 2}, World Scientific,
Singapore (1987).}
\bref\Lu{
G. Lusztig, J. Am. Math. Soc. {\bf 3} (1990) 447; Geom. Dedic. {\bf 35}
(1990) 89.}
\bref\LZ{
B.H. Lian and G.J. Zuckerman, Phys. Lett. {\bf 254B} (1991) 417; preprint
YCTP-P18-91.}
\bref\BMPgrav{
P. Bouwknegt, J. McCarthy, and K. Pilch, preprint CERN-TH.6162/91.}
\bref\BMPprep{
P. Bouwknegt, J. McCarthy, and K. Pilch, in preparation.}

\newsec{ Introduction}

The problem we address in this letter arises quite naturally in
the so-called free field approach to conformal field theories.  The
mathematical question one wants to answer within this framework is
whether a given irreducible module $\cal L$ of the chiral algebra $\cal
A$ has a resolution in terms of Fock spaces.  More precisely, one wants
to construct a family of free field Fock spaces $\cF^{(i)}$, which are
$\cal A$-modules, and a set of $\cal A$-homomorphisms (intertwiners)
$d^{(i)}:\cF^{(i)}\to \cF^{(i+1)}$, such that these spaces with the
maps between them form a complex whose (co)homology is isomorphic to
$\cal L$. The first example of such a construction was given by Felder
[\Fe] for the class of representations of the Virasoro algebra
corresponding to minimal models [\BPZ]. Later, this was
extended to other conformal field theories, in particular WZNW  and
their coset models (for review see [\BMPrev] and references therein).

Felder's construction relies on the complete classification of
submodules of Fock spaces given by Feigin and Fuchs [\FFear,\FF], and the
explicit form of the intertwiners. The latter are defined, following
Thorn [\Thorn], as multiple integrals of products of the screening currents.
One may wonder whether the restriction to the
representations in the fundamental range of the minimal series is
important. The answer turns out to be negative, and already examples
of similar resolutions outside this series have been discussed
in [\FFK].  In the following we will argue that a Fock space resolution
can be explicitly constructed for any irreducible highest weight
module of the Virasoro algebra provided one introduces additional
intertwiners besides those considered in [\Fe,\FFK]. The existence
of such intertwiners, and their properties needed in
the computation of the cohomology, were demonstrated by Tsuchiya and
Kanie [\TK]. We will discuss some of their results within the
Dotsenko-Fateev formalism [\FD], as used in [\Fe,\BMPrev],
which may be more familiar. Not to make our presentation too long
we will restrict to modules with $c\leq 1$.

Our three main results, which cover the cases not analyzed previously, are
summarized in Theorems 5.2, 6.1 and 7.1. In particular, the first gives the
resolution for modules corresponding to the so-called boundary of
the Kac table [\Ka], whilst the last extends Felder's construction to
modules outside the fundamental range.

This letter is organized as follows: In Section 2, as well as
introducing some definitions and notation, we summarize the results of
Feigin and Fuchs which will be used later. Then, in Sections 3 and 4, we
introduce intertwiners and recall the construction of Felder's complex.
After this review, we discuss the boundary case in detail in
Sections 5 and 6,
and in Section 7 describe the extension of the resolution for the
irreducible modules beyond the fundamental range. We conclude in Section 8
with some remarks on possible applications of these results.
\vfill\eject

\newsec{Feigin-Fuchs modules}

The generators of  the Virasoro algebra, $\it Vir$,
satisfy\footnote{$^*$}{See,\eg [\GSW] for basic definitions and  [\BPZ,\FMS]
for a review of conformal field theory techniques used throughout this
letter.}
\eqn\KPaa{[L_m,L_n]=(m-n)L_{m+n} + {c\over 12}m(m^2-1)\de_{m+n,0}\,, \quad
[L_n,c]=0\,,\qquad m,n\in\ZZ\,.}
In this letter we will consider three classes of highest weight modules
of $\it Vir$: the irreducible modules $\cL_{h,c}$, the Verma modules
$\cM_{h,c}$, and the Feigin-Fuchs modules $\cF_{p,Q}$. We recall that
the central charge, $c$, and the conformal dimension, $h$, determine
$\cL_{h,c}$ and $\cM_{h,c}$ completely, and that the latter is
generated freely by $L_{-n}$, $n\geq 1$, acting on the vacuum $v_{h,c}$,
$L_0 v_{h,c}=h v_{h,c}$.

The Feigin-Fuchs module $\cF_{p,Q}$
is just the Fock space of a  scalar field, $\phi(z)$, with a background
charge $Q$ and momentum $p$ such that
\eqn\KPab
{c= 1-12Q^2\,,\quad h=\half p(p-2Q)\,.}
We follow  the convention that the two-point function of  $\ph(z)$ is
\eqn\LZac{\vev{ \ph(z)\ph(w) }  = - \ln (z-w)\,,}
and the stress energy tensor, $T(z)=\sum_{n\in\ZZ}L_nz^{-n-2}$, has the form
\eqn\LZab{T(z) = -\half: \p\ph(z)\p\ph(z): + iQ \p^2\ph(z)\,.}
In terms of modes  $i\p\ph(z) = \sum_{n\in\ZZ} \al_nz^{-n-1}$,
\eqn\LZcx{L_n = \half \sum_{m\in\ZZ} :\al_m\al_{n-m}:  -  (n+1)Q \al_n\,,}
where
\eqn\LZd{[\al_m,\al_n] = m\de_{m+n,0}\,.}
The Fock space $\cF_{p,Q}$ is generated from the vacuum $v_{p}$,
$\al_0 v_{p}=pv_{p}$, by the free action of the creation operators
$\al_{-m}$, $m\ge 1$.

As can easily be seen from \LZcx, $\cF_{p,Q}$ and $\cF_{-p,-Q}$ are
isomorphic as Virasoro modules. Choosing one solution for $Q$
in $\KPab$ leaves us, for a given $h$ and $c$, with two Fock spaces
$\cF_{p,Q}$ and $\cF_{2Q-p,Q}$, dual to each other,  \ie $\cF_{p,Q}^*\simeq
\cF_{2Q-p,Q}$.

The detailed structure of submodules of all  Verma and Feigin-Fuchs
modules of the Virasoro algebra has been obtained in [\FFear,\FF].
There are three main types of modules, $I$, $II$ and $III$,
depending on whether the equation
\eqn\KPba{x\al_++y\al_-+\sqrt{2}(p-Q)=0\,,}
has respectively zero, one or infinitely many
integral solutions for $x$ and $y$. Here we have introduced
$\al_+$ and $\al_-$, $\al_+\al_-=-1$,
which also parametrize the background charge,
\eqn\KPac{Q=\sqrt\half (\al_++\al_-) \,. }

In the first case (type $I$) the Verma module and the Fock space are
isomorphic and irreducible [\FF], and obviously no resolution is
needed. Thus we will assume that  \KPba\ has at least one integral
solution, or, equivalently, that the momentum $p$ can be parametrized
by a pair of integers $n$ and $n'$ (\ie there is exactly one such pair
for type $II$, and for modules of type $III$ infinitely many such),
\eqn\KPbb{p=\pp{n,n'}=\sqrt\half \left((1-n)\al_++(1-n')\al_-\right)\,.}
For $\pp{n,n'}$ as in \KPbb\ we denote the corresponding conformal
dimension, computed from \KPab , by $\hh{n,n'}$,
and, for a fixed $Q$ (and $c$) write $\cF_{n,n'}$  and $\cM_{n,n'}$
instead of $\cF_{\pp{n,n'},Q}$ and $\cM_{\hh{n,n'},c}$.  Virasoro
modules with the momentum given in \KPbb\ arise in the
generalized Dotsenko-Fateev minimal models [\FD,\FFK].

In case $III$, in order to have infinitely many integral solutions to
\KPba, there must exist (relatively prime) integers $p$, $p'$,
$pp'\not=0$,  such that $p\al_++p'\al_-=0$. We can then take
$\al_+=\sqrt{p'/p}$ and $ \al_-= -\sqrt{p/p'}$. Clearly the central
charge must be rational, and, depending on whether $pp'> 0$ or
$pp'< 0$ we have $c\leq 1$ or $c\geq 25$, respectively. In the
following we will restrict ourselves to $c\leq 1$ and take $p'\geq p
\geq 1$. In particular $c=1$ corresponds to $p'=p=1$, and for $c<1$ we
must have $p'>p$ .  Using the freedom in the parametrization of the
momentum, $\pp{n,n'}= \pp{n+jp,n'+jp'}$, $j\in\ZZ$, we can always set
$0\leq n'\leq p'-1$ (or, equivalently, $0\leq n\leq p-1$).
If, in addition, $p'>p>1$ and
\eqn\KPaf{1\leq n\leq p-1\,, \quad 1\leq n'\leq p'-1\,,}
we will say that the module belongs to the fundamental range of
the minimal series [\BPZ], or to the interior of the Kac table [\Ka].
The exterior of the table is defined by letting $n\not=0\,\,\mod{p}$
to be outside the range \KPaf . Finally,  $n=0\,\,\mod p$ or $n'=0
\,\,\mod{p'}$ correspond to the boundary of the table.

In cases $II$ and $III$,  the detailed structure of
$\cF_{n,n'}$ as a Virasoro module depends on
the momentum $\pp{n,n'}$ \KPbb , and all possible subcases (in case
$III$ with $c\leq 1$) are listed in the following theorem which
summarizes the results of Feigin and Fuchs.\thm\KPTff
\def\bul{$\bullet$}

\proclaim  Theorem \KPTff. ([\FF]) In case $II$ there are four subcases:
\item{\bul} Case  $II_-$ (respectively $II_0$): If $nn'<0$
(respectively $nn'=0$) then
$\cF_{{n,n'}}$ and $\cM_{{n,n'}}$ are isomorphic and irreducible.
\item{\bul} Case $II_+(-)$: If $n,n'<0$ then
$\cF_{{n,n'}}\simeq\cM_{{n,n'}}$ are reducible. The maximal
submodule is isomorphic with $\cF_{{-n,n'}}\simeq\cM_{{-n,n'}}$.
\item{\bul} Case $II_+(+)$: This case, where $n,n'>0$, is dual to $II_+(-)$,
\eg $\cF_{n,n'}\simeq\cM_{-n,-n'}^*$.
\smallskip
\noindent
In case $III$ for $c\leq 1$ there are four
subcases, in all of which the Fock space is a reducible module.
In the first three, $III_-$ and $III_-^0(\pm)$ below, $p'>p>1$ (\ie $c<1$),
and $m,\,m'$  denote labels in the fundamental range, $1\leq m\leq p-1$,
$1\leq m'\leq p'-1$.
In $III_-^{00}$ we have $p'\geq p\geq 1$ (\ie $c\leq 1$). In all
cases $j\in\ZZ$.
\item{\bul} Case $III_-$: The submodules of $\cF_{m+jp,m'}$
are generated by the vectors $u_i$, $v_i$, and $w_i$ given by the following
diagram
\def\cs{\,{\smash{\mathop{\swarrow\llap{$\hbox{$\searrow\,\,$} $} }}}}
 \def\se{\searrow}
\eqn\KPbc{{\matrix{
v_0&\ot&w_0&\to&v_1   &\ot&w_1&\to&v_2   &\ot&w_2&\to&\cdots  \cr
   &\se&   &\cs&      &\cs&   &\cs&      &\cs&   &\cs&\cdots  \cr
   &   &u_1&\ot&v_{-1}&\to&u_2&\ot&v_{-2}&\to&u_3&\ot&\cdots  \cr}\quad,}}
whose conformal weights are (see [\Fe])
\eqn\KPbe{\eqalign{
h(v_0)&=\hh{m+jp,m'}\,,\cr
h(u_{i})&=\hh{-m+(|j|+2i)p,m'}\,,\,\,i\geq 1\,,\cr
h(v_{-i})&=\hh{m-(|j|+2i)p,m'}\,,\,\,i\geq 1\,,\cr}
\qquad \eqalign{&\cr
h(w_{i})&=\hh{-m-(|j|+2i)p,m'}\,,\,\,i\geq 0\,,\cr
h(v_{i})&=\hh{m+(|j|+2i)p,m'}\,,\,\,i\geq 1\,.\cr}}
\item{\bul}  Case $III_-^0(-)$: The submodules of
i) $\cF_{m+jp,0}$, $j\geq 0$, and ii) $\cF_{0,m'+jp'}$, $j\geq 0$, are
generated by the vectors $u_i$ and $v_i$,
\eqn\KPaj{\matrix{
v_0&\to&u_1&\ot&v_1&\to&u_2&\ot&v_2&\to&u_3&\ot&\cdots\cr}\,,}
where
\eqn\KPbj{\eqalign{{\sl i)}&\cr {\sl ii)}&\cr}\quad
\eqalign{
h(v_i)&=\hh{m+(j+2i)p,0}\,,\,\, i\geq 0\,,\cr
h(v_i)&=\hh{0,m'+(j+2i)p'}\,,\,\, i\geq 0\,,\cr}
\quad
\eqalign{
h(u_i)&=\hh{m-(j+2i)p,0}\,,\,\, i\geq 1\,;\cr
h(u_i)&=\hh{0,m'-(j+2i)p'}\,,\,\, i\geq 1\,.\cr}}
\item{\bul}  Case $III_-^0(+)$: The submodules of
i) $\cF_{m+jp,0}$, $j< 0$,  and ii) $\cF_{0,m'+jp'}$, $j< 0$, are
generated by the vectors $u_i$ and $v_i$,
\eqn\KPai{\matrix{
u_1&\ot&v_1&\to&u_2&\ot&v_2&\to&u_3&\ot&v_3&\to&\cdots\cr}
\,.}
where
\eqn\KPbi{\eqalign{{\sl i)}&\cr {\sl ii)}&\cr}\quad
\eqalign{
h(v_i)&=\hh{m-(j-2(i-1))p,0}\,,\,\, i\geq 1\,,\cr
h(v_i)&=\hh{0,m'-(j-2(i-1))p'}\,,\,\, i\geq 1\,,\cr}
\quad
\eqalign{
h(u_i)&=\hh{m+(j-2(i-1))p,0}\,,\,\, i\geq 1\,;\cr
h(u_i)&=\hh{0,m'+(j-2(i-1))p'}\,,\,\, i\geq 1\,.\cr}}
\item{\bul} Case $III_-^{00}$: The Fock space $\cF_{jp,0}\simeq
\cF_{-jp,0}$ is the direct sum of
irreducible modules, $\cF_{jp,0}=\bigoplus_{k=0}^\infty \cL_{(|j|+2k)p,0}$.

In the diagrams \KPbc, \KPaj, and \KPai\ vectors $u_i$ correspond to
singular vectors in the Fock space $\cF_{\star,\star}$ and generate the
submodule,  $\cF_{\star,\star}'$, which is a direct sum of irreducible
highest weight modules.  In the quotient  $\cF_{\star,\star}/
\cF_{\star,\star}'$ vectors $v_i$ become singular, and generate
$\cF_{\star,\star}''$ which is a direct sum of irreducible highest
weight modules. Note that $v_0$ is special, since it is singular.
Finally, $w_i$ are singular in $\cF_{\star,\star}'/
\cF_{\star,\star}''$, and generate a direct sum of irreducible highest
weight modules.  The arrows, $1\to 2$, indicate that the second vector
is in the submodule generated by the first one.

There are also similar diagrams for the composition series of singular
vectors of the Verma modules, in which the singular vectors occur at
precisely the same conformal weights as above. Moreover, some arrows
must be reversed, so that all of them describe embeddings of the Verma
submodules.

\newsec{The intertwiners}

\def\smpl{{\scriptscriptstyle (+)}}
\def\smmi{{\scriptscriptstyle (-)}}
\def\tooo{\longrightarrow}

Let us first introduce a class of intertwiners between Fock spaces.
They are constructed as products of the screening currents
\eqn\KPae{s_+(z)=:\exp(i\sqrt 2\al_+\phi)(z):\,\quad {\rm and }\quad
s_-(z)=:\exp(i\sqrt 2\al_-\phi)(z):\,,}
integrated over suitable multiple contours [\Thorn,\FF,\TK,\Fe].
For positive integers $r$ and $r'$ let us consider operators
\eqn\KPinter{
Q_{r}^\smpl[\Om]=\con{(s_+)^r}_{\Om_r}\,,\quad
Q_{r'}^\smmi[\Om]=\con{(s_-)^{r'}}_{\Om_{r'}}\,,}
where $\con{\cdots}_{\Omega_k}$ is defined by
\eqn\KPbh{
\con{s_{i_1}\ldots s_{i_k}}_{\Omega_k}=\int_{\Omega_k}
dz_1 \ldots dz_{k} \,  s_{i_1}(z_1)\ldots s_{i_k}(z_{k})\,,\quad i_1,\ldots,
i_k\in\{+,-\}\,.}
and $\Om_k$ is a set of contours for $z_1,\ldots,z_{k}$.
When acting on $\cF_{n,n'}$, the integrand in \KPbh\ is analytic in
$M_k=\CC^k\setminus \{z_i=0,z_i=z_j;i,j=1,\ldots,k\}$  and has
nontrivial monodromies due to the presence of the factor
\eqn\KPcb{\prod_{i<j}(z_i-z_j)^{2\al_\pm^2}
\prod_{k}z_k^{\sqrt 2 \al_\pm \pp{n,n'}}\,.}
In order that \KPbh\ be a well-defined operator we must require
that $\Om_k$ is a closed contour [\TK], \ie it is an element of
$H(M_k,{\cal S}_{\al_\pm^2})$, the homology  of $M_k$ with coefficients in
the local system ${\cal S}_{\al_\pm^2}$ corresponding to \KPcb\ (this
system depends only on $\al_\pm^2$).

In explicit computations one needs to have convenient representatives
of these homology classes.  We will construct them using the following two
classes of multicontours (singular at a point) that have been used
in [\Fe,\BMPcmp,\BMPrev]:
In the first class, which we denote $\Ga_k$, the integration
variables $z_1,\ldots,z_{k}$ are
taken counterclockwise from 1 to 1 around 0, and nested according to
$|z_1|>\ldots > |z_k|$. In the second, $\widehat\Ga_k$,
the $z_1$ integration is along a
contour surrounding 0, while $z_2,\ldots z_k$ are integrated
counterclockwise from the base point $z_1$ to $z_1$ around 0, and the
nesting is the same as in $\Ga_k$.  The ambiguity in the phase
of the integrand \KPbh\ is fixed by analytic continuation from the
positive real half-line (see [\Fe,\BMPcmp]). Let us  denote the
resulting  operators \KPinter\ by $Q_r^\smpl$, $Q_{r'}^\smmi$ and
$\widehat Q_r^\smpl$, $\widehat Q_{r'}^\smmi$, respectively.

We will also consider multiple contours obtained by putting
several of those together. For example $\Ga_k\cup\Ga_{k'}$
will denote a multiple contour $z_1,\ldots z_k,\ldots z_{k+k'}$
in which the variables of $\Ga_{k'}$ are nested inside those of $\Ga_k$.

We may now state  the main result about the intertwiners between
Feigin-Fuchs modules of type $III$ due
to Tsuchiya and Kanie [\TK]. \thm\KPTinter

\proclaim Theorem \KPTinter. 1)  Consider $\cF_{n,n'}$, where
 $n=m+jp$, $1\leq m\leq p$, $j\in \ZZ$,  and $n'$ is arbitrary. Then
for any non-negative integer $k$ the operator
\eqn\intermap{
Q_{m+kp}^\smpl[\Om]:\cF_{n,n'}\tooo\cF_{n-2(m+kp),n'}\,,}
is a well defined intertwiner provided $\Om_{m+kp}\in
H(M_{m+kp},{\cal S}_{\al_+^2})$.\smallskip\noindent
2) These intertwiners are nontrivial in the sense that such
 $\Om_{m+kp}$ exists for any $k\geq 0$ and \smallskip
\item{i)} if $n'+(k-j)p'\leq 0$ then $Q_{m+kp}^\smpl[\Om]v_{n,n'}\not=0$ ;
\item{ii)} if $n'+(k-j)p'> 0$  then $Q_{m+kp}^\smpl[\Om]\chi=
v_{n-2(m+kp),n'}$ for some $\chi\in\cF_{n,n'}$ .\smallskip\noindent
Here $v_{n,n'}$ and $v_{n-2(m+kp),n'}$ denote vacua of
$\cF_{n,n'}$ and $\cF_{n-2(m+kp),n'}$, respectively.\smallskip
The analogous result holds for the operators $Q^\smmi_\star[\star]$.

\newsec{The Fock space resolution for type $II_+$ and $III_-$ modules  }

We begin by constructing explicitly  intertwiners using
contours $\Ga_r$, $\Ga_{r'}$,  $\widehat \Ga_r$ and
$\widehat \Ga_{r'}$, $1\leq r\leq p$ and $1\leq r'\leq p'$, introduced
in the previous section. \thm\KPTph

\proclaim Lemma \KPTph.
Operators  $Q_r^\smpl\,,\,\widehat Q_r^\smpl:
\cF_{n,n'}\tooo \cF_{n-2r,n'}$ and  $Q_{r'}^\smmi\,,\,\widehat Q_{r'}^\smmi:
\cF_{n,n'}\tooo \cF_{n,n'-2r'}$
are well defined intertwiners between Virasoro modules if
$r=n\,\mod{p}$ and $r'=n'\,\,\mod{p'}$, respectively.

\proof   Operators $\widehat Q_r^\smpl$ and $\widehat Q_{r'}^\smmi$
are well defined and commute with the action of the Virasoro algebra
 iff $\widehat \Ga$ is a closed cycle.  This amounts to the
$z_1$-contour being closed.  For the values of $r$ and $r'$ as in the
lemma this is easily verified using standard  methods [\Fe].
 For $Q_r^\smpl$ and $Q_{r'}^\smmi$ we use the following lemma which
can be proven by standard manipulations with the contours
[\Fe,\BMPcmp]. \hfill\Box

\thm\KPTzer
\proclaim Lemma \KPTzer. For $1\leq r\leq p$ and $1\leq r'\leq p'$,
\eqn\KPcc{Q_r^\smpl={1\over r}{1-q_+^{2r}\over 1-q_+^2}\widehat
Q_r^\smpl\,,\quad
Q_{r'}^\smmi={1\over r'}{1-q_-^{2r'}\over 1-q_-^2} \widehat Q_{r'}^\smmi\,,}
where $q_\pm=\exp(i\pi\al_\pm^2)$.

Note that for positive $r_1$, $r_2$, $r_1+r_2\leq p$ we have
$Q_{r_1}^\smpl Q_{r_2}^\smpl =Q_{r_1+r_2}^\smpl$, provided
$Q_{r_1}^\smpl$ and $ Q_{r_2}^\smpl$ act on spaces on which they are
well-defined. In particular -- after further investigation of the integral
in $\widehat Q_p^\smpl$ --
\KPcc\ implies
\eqn\KPcd{Q_m^\smpl Q_{p-m}^\smpl = Q_{p-m}^\smpl Q_m^\smpl
= Q_p^\smpl=0\,,}
The same also holds for $Q_{m'}^\smmi$.

Lemma \KPTph\ together with  identity \KPcd\ is the basis for the construction
of the complex of Fock spaces when $m$, $m'$ lie in the fundamental range
\KPaf . This complex,
$(\cF,d)\equiv \{(\cF^{(i)}_{m,m'},d^{(i)}),i\in\ZZ\}$, is defined as follows
\eqn\KPce{
\eqalign{\cF^{(2j)}_{m,m'}&=\cF_{m-2jp,m'}\,,\cr
         d^{(2j)}&=Q_{m}^\smpl\,,\cr} \quad
\eqalign{\cF^{(2j+1)}_{m,m'}&=\cF_{-m-2jp,m'}\,,\cr
         d^{(2j+1)}&=Q_{p-m}^\smpl\,,\cr}\qquad j\in\ZZ\,.}

Using the submodule structure of Fock spaces $\cF^{(i)}_{m,m'}$
as summarized in Theorem \KPTff ,  Felder [\Fe] was able to compute
the kernels and the images of all the intertwiners $d^{(i)}$, and
prove the following important result \thm\KPTfeld

\proclaim Theorem \KPTfeld. Let $1\leq  m\leq p-1$, $1\leq m'\leq
p'-1$. Then the complex $(\cF,d)$  defined in \KPce\ is a (two-sided)
resolution of the irreducible module $\cL_{m,m'}$, \ie
\eqn\KPce{H^{(i)}(\cF,d)\simeq\de_{i,0}\cL_{m,m'}\,.}

In fact there are three more resolutions, one in terms  of modules
$\cF_{p-m,p'-m'}^{(i)}$ (recall that $\cF^*_{m,m'}
\simeq\cF_{p-m,p'-m'}$ and $\cL_{p-m,p'-m'}\simeq \cL_{m,m'}$), and two
others if we use operators $Q_\star^\smmi$ instead of $Q_\star^\smpl$.

It may be worth noting that  Felder's proof also shows that the
differential in the complex is nilpotent without resorting to  the
computation in  \KPcd.

A similar construction for the modules of type $II_+(\pm)$ was carried out
in [\FFK]. \thm\KPTffk

\proclaim Theorem \KPTffk. In the notation of Theorem \KPTff, let
$n,n'<0$. Then the complex
\eqn\KPcf{
\matrix{0& \tooo & \cF_{-n,n'}& \mapright{ Q_{-n}^\smpl} &
\cF_{n,n'} & \tooo & 0 \,,\cr}}
is a resolution for $\cL_{n,n'}$. For the case $II_+(+)$  the
resolution can be obtained by dualization of  \KPcf .

\newsec{ Resolutions for the  modules of type $III_-^0$}

Using the results reviewed in the previous sections we will now
construct resolutions for irreducible modules $\cL_{h,c}$ of type
$III_-^0$, for which $h=h_{m+jp,0}$, $1\leq m\leq p-1$  or
$h=h_{0,m'+jp'}$, $1\leq m'\leq p'-1$, $j\in\ZZ$.
Since $h_{n,n'}=h_{-n,-n'}$, we may take $j\geq 0$.

Let us consider the Fock space $\cF_{0,m'+jp'}$.  According to Theorem
\KPTff\ this space has type $III_-^0(-)$ and its maximal submodule,
$\cF'_{0,m'+jp'}$, is generated by vectors $u_i$, $v_i$, $i\geq 1$. It
also follows from \KPaj\ that the irreducible module is isomorphic to
the quotient, namely
\eqn\KPfa{\cL_{0,m'+jp'}=\cF_{0,m'+jp'}/\cF'_{0,m'+jp'}\,.}
By comparing \KPbj\ and \KPai\ we find that the weights of $u_i$ and
$v_i$ in $\cF'_{0,m'+jp'}$ are precisely those of the vectors generating
submodules of the Fock space $\cF_{0,m'-(j+2)p'}$. Let us denote the
latter vectors by $u_i'$, $v_i'$, $i\geq 1$. This observation and
\KPfa\ reduce the problem of constructing a resolution to that of
finding a homomorphism $d$ from $\cF_{0,m'-(j+2)p'}$ into $\cF_{0,m'+jp'}$,
such that $du'_i=u_i$ and $dv_i'=v_i$, $i\geq 1$. Note that then $d$
must be injective, \ie it is an embedding.

In a sense the problem is similar to the one in case $II_+(-)$, except
that the structure of submodules of Fock spaces is somewhat more
complicated.  Since $\cF_{0,m'+jp'}\simeq \cF_{-jp,m'}$ and
$\cF_{0,m'-(j+2)p'} \simeq \cF_{(j+2)p,m'}$, the intertwiner
should involve  $(j+1)p$ currents $s_+(z)$. By Theorem \KPTinter\
we know it exists. To construct it explicitly we observe that,
in view of Lemma
\KPTph\ and \KPcd , an obvious building block for such operators is
$\widehat Q_{p}^\smpl$.  Its properties are summarized in the
following technical lemma whose proof will be outlined at the end of
the section. \thm\KPTtech
\def\mapp{\widehat Q_{p}^\smpl}
\def\do{\downarrow}

\proclaim Lemma \KPTtech. Consider $\mapp:\cF_{0,m'+kp'}\tooo
\cF_{0,m'+(k+2)p'}$, $k\in\ZZ$. Depending on $k$ there are four
cases described by the diagrams i)--iv) below. Operator
$\mapp$ maps special vectors $u_i'$ and $v_i'$ from the first Fock space
onto $u_i$ and $v_i$ in the second space, as indicated
by the downward arrows, \ie $\mapp$ is nonzero along these arrows. \smallskip
$$\eqalignno{
&\matrix{{\sl i)}& k\leq -3\,:&&
   &   &    &   &u_1'&\ot&v_1'&\to&u_2'&\ot&\cdots\cr
 &&&          &   &    &   &\do &   &\do &   &\do &   &      \cr
 &&&       u_1&\ot&v_1 &\to&u_2 &\ot&v_2 &\to&u_3 &\ot&\cdots\cr}&\cr
\noalign{\vskip10pt}
&\matrix{ {\sl ii)}& k=-2 \,:&&
  &   &u_1'&\ot&v_1'&\to&u_2'&\ot&v_2'&\to&\cdots\cr
 &&&        &   &\do &   &\do &   &\do &   &\do &   &      \cr
 &&&    v_0 &\to&u_1 &\ot&v_1 &\to&u_2 &\ot&v_2 &\to&\cdots\cr}&\cr
\noalign{\vskip10pt}
&\matrix{{\sl iii)}& k=-1\,:&&
  u_1'&\ot&v_1'&\to&u_2'&\ot&v_2'&\to&u_3'&\ot&\cdots\cr
 &&&           &   &\do &   &\do &   &\do &   &\do &   &      \cr
 &&&           &   &v_0 &\to&u_1 &\ot&v_1 &\to&u_2 &\ot&\cdots\cr}&\cr
\noalign{\vskip10pt}
&\matrix{{\sl iv)}&k\geq 0\,:&&
  v_0'&\to&u_1'&\ot&v_1'&\to&u_2'&\ot&v_2'&\to&\cdots\cr
 &&&           &   &    &   &\do &   &\do &   &\do &   &      \cr
 &&&           &   &    &   &v_0 &\to&u_1 &\ot&v_1 &\to&\cdots\cr}&\cr
}$$

The main result of this section is \thm\KPTres

\proclaim Theorem \KPTres.  The intertwiner  $d=(\widehat
Q_{p}^\smpl)^{j+1}$ is an embedding of $\cF_{0,m'-(j+2)p'}$ into
$\cF_{0,m'+jp'}$, \ie the complex
\eqn\KPfb{\matrix{0&\tooo&\cF_{0,m'-(j+2)p'}&\mapright{d}&\cF_{0,m'+jp'}&
\tooo&0\cr}\,,} is a Fock space resolution of $\cL_{0,m'+jp'}$. Another
resolution is given by the dual complex
\eqn\KPfc{\matrix{0&\tooo&\cF_{0,-m'-jp'}&\mapright{d^*}&\cF_{0,-m'+(j+2)p'}&
\tooo&0\cr}\,,} where $d^*=(\widehat Q_{p}^\smpl)^{j+1}$.

\proof Using Lemma \KPTtech\ we can compute the image of
$\cF_{0,m'-(j+2)p'}$ when acting with subsequent $\mapp$. The result is
$d(\cF_{0,m'-(j+2)p'})=\cF'_{0,m'+jp'}$. The second part of the theorem
follows from $\cL_{0,m'+jp'}^*\simeq \cL_{0,m'+jp'}$ and
$\mapp{}^*=\mapp$.\hfill\Box

Clearly, for modules $\cL_{m+jp,0}$ there are analogous resolutions in
which the differential is $\widehat Q_{p'}^\smmi$.
\smallskip

\noindent
{\it Proof of Lemma \KPTtech:} The general idea of the proof is the same
as that of parts of Theorem \KPTfeld\ (see [\Fe]).
Let us begin with case i). By Theorem \KPTinter . 2.i)  $\mapp u_1'\not=0$.
To verify this we choose a covector $\chi\in\cF_{0,m'+(k+2)p'}^*$
(see [\Thorn] and Appendix in [\Fe])
such that
\eqn\KPfz{
\langle\chi,s_+(z_1)\ldots s_+(z_{p})u_1'\rangle=
\prod_{\ell=1}^p z_\ell^{-(k+1)p'-m'}
\prod_{\ell,\ell'=1\atop \ell<\ell'}^{p}
(z_\ell-z_{\ell'})^{2\al_+^2}
\prod_{\ell=1}^p z_\ell^{\sqrt 2 \al_+ \pp{0,m'+kp'}}\,.}
Then a straightforward computation yields
\eqn\KPfx{\eqalign{
\langle\chi,\mapp u_1'\rangle&=
\int_{\widehat \Ga}dz_1\ldots dz_{p}\,
\prod_{\ell,\ell'=1\atop \ell<\ell'}^{p}
(z_\ell-z_{\ell'})^{2\al_+^2} \prod_{\ell=1}^{p} z_\ell^{\al_+^2-p'-1}\cr
&=2\pi i (-1)^{p-1}\prod_{\ell=1}^{p-1}{(1-q_+^{2\ell})^2\over
1-q_+^2} {\cal J}_{0,p-1}(2\al_+^2,\al_+^2-p'-1,\al_+^2)\cr
&=(2\pi i)^{p} (-1)^{pp'+p-1} {p'\,!\over (p-1)\,!}
{e^{i\pi\al_+^2}
\over \Ga(1+\al_+^2)^{p} } \prod_{\ell=1}^{p-1} {\sin\pi\ell\al_+^2
\over\sin\pi\al_+^2}\cr
&\not=0\,.\cr}}
We used here an explicit result for the Dotsenko-Fateev integral [\FD]
\eqn\KPfv{\eqalign{
{\cal J}_{0,n}(\al,\be;\rho)&\equiv {1\over n\,!}\int_0^1dt_1\ldots dt_n
\prod_{\ell=1}^n (1-t_\ell)^\al t_\ell^\be
\prod_{\ell,\ell'=1\atop\ell<\ell'}^n |t_\ell-t_{\ell'}|^{2\rho}\cr
&=\prod_{\ell=1}^n {\Ga(\ell\rho)\over\Ga(\rho)}{\Ga(1+\al+(\ell-1)\rho)
\Ga(1+\be+(\ell-1)\rho)\over\Ga(2+\al+\be+(n+\ell-2)\rho)}\,.\cr}}

The rest of the lemma in this case follows entirely from the embedding
patterns of submodules in both Fock spaces and  $\mapp$ being  an
intertwiner. Let us outline the first few steps.

Normalize $u_2$ such that $u_2=\mapp u_1'$. Since $u_1'$ is in the
submodule generated by $v_1'$ we must have $\mapp v_1'\not=0$.  Note
that the latter remains nonzero after we divide $\cF_{0,m'+(k+2)p'}$ by
the submodule generated by $u_1$ and $u_2$. Indeed, suppose the
opposite, \ie that  $\mapp v_1'=P_1u_1+P_2u_2$, where $P_{1,2}$ are
some polynomials in $L_{-n}$, $n\geq 1$. Since the submodule generated
by $u_1$ and $u_2$ is a direct sum of two irreducible ones, we also
have $u_1=P'_1P_1u_1$ where $P'_{1}$ is some element in the enveloping
algebra of $\it Vir$. But if $P_1u_1\not=0$ then
$u_1=\mapp P_1'(v_1'-P_2u_1')$, and a simple examinination of the weights
in both Fock spaces shows the r.h.s.\ must vanish, which is
a contradiction. Thus we can at most have $\mapp v_1'=P_2u_2$. But then
$\mapp(v_1'-P_2u_1')=0$, \ie $ v_1'-P_2u_1'\not=0$ is in the kernel of
$\mapp$. Since the submodule $ker\,\mapp$ contains neither
$u_1'$ nor $v_1'$,  there can be no such vectors at this level, \ie we
must have $\mapp v_1'\not= P_2u_1$. That  $\mapp v_1'$ is singular
after we divide out the submodule generated by $u_1$
follows from the corresponding property for $v_1'$.
Thus, finally,
we may set $\mapp v_1'=v_2$.  In the next step a similar
reasoning shows that $\mapp u_2'\not=0$, so, up to normalization it
yields $\mapp u_2'=u_3$. And so on.

Case ii) is proven by exactly the same method. Cases iii) and iv) are
dual to ii) and i), respectively. Since $\mapp{}^*=\mapp$, we
deduce from  $coker\,\mapp{}^*\simeq ker\,\mapp$  that $\mapp$ must be
{\it onto}.  Because $\mapp$ commutes with {\it Vir}, $u'_i$ are mapped
onto $u_i$, and the maps indicated by the arrows are clearly nonzero.
After we divide  both sides by  submodules generated by $u_i'$ and
$u_i$, respectively, $\mapp$ becomes a homomorphism from a direct sum
of irreducible highest weight modules onto a direct sum of a subset of
these modules, and the arrows between $v_i'$ and $v_i$ follow. \hfill\Box

\newsec{Resolutions for the modules of type $III_-^{00}$}

The problem of obtaining a resolution for a  module $\cL_{jp,0}$,
$j\geq 0$, is similar to the one discussed above. Given
the Fock space $\cF_{jp,0}$ we must construct an embedding
from $\cF_{(j+2)p,0}$ into $\cF_{jp,0}$. The result is \thm\KPTcone

\proclaim Theorem \KPTcone. Let $p'\geq p\geq 1$ and $j\geq 0$.
The operator $\mapp :\cF_{(j+2)p,0}\tooo \cF_{jp,0}$ is an embedding
and $\cL_{jp,0}\simeq \cF_{jp,0}/\mapp(\cF_{(j+2)p,0})$.

\proof Since $\cF_{jp,0}$ and $\cF_{(j+2)p,0}$ are direct sums of
irreducible highest weight modules, the proof essentially consists of
verifying that $\mapp$ does not annihilate any of the singular vectors
in $\cF_{(j+2)p,0}$. For simplicity let us only consider the case
$p=p'=1$, \ie $c=1$. Then $\widehat Q_1^\smpl=\oint dz :\exp(i\sqrt2
\phi(z)):$ is just a vertex operator. Using explicit formulae (in terms
of Schur polynomials) for
singular vectors in the Fock space $\cF_{kp,0}$, $k\geq 0$ [\WY,\KR],
it is easy to check that, up to normalization,  they coincide with the
vectors $(\widehat Q_1^\smpl)^\ell v_{kp+2\ell,0}$, $\ell \geq 0$,
where $v_{kp+2\ell,0}$ denotes the vacuum of $\cF_{kp+2\ell,0}$.
The theorem is then obvious. The general case is similar. \hfill\Box

\newsec{Generalizations of Felder's construction}

Consider an irreducible module, $\cL_{m+jp,m'}$, of type $III_-$, where
$m,m'$ lie in the fundamental range \KPaf\ and $j$ is an arbitrary
integer.
The Fock space $\cF_{m+jp,m'}$ appears in the complex $(\cF,d)$ corresponding
to the resolution of $\cL_{m,m'}$ for $j$ even, and $\cL_{p-m,m'}$ for
$j$ odd, constructed as in Section 4.  An important feature of $(\cF,d)$
is that the vacuum of the 0-th Fock space has the highest weight
with respect to all other states in the complex. In particular, this
guarantees that the cohomology  contains at least the irreducible
module. Let us now consider the collection of Fock spaces obtained by
removing from $(\cF,d)$ all Fock spaces whose
weights are higher than $h_{m+jp,m'}$; \ie we delete the spaces
$\cF_{-m+2jp,m'},\ldots,\cF_{m-2jp,m'}$ for $j>0$, and the spaces
$\cF_{m-jp,m'},\ldots, \cF_{-m+(j+2)p,m'}$ for $j<0$.
On both sides of the deleted segment the differential is  defined as
before in terms of $Q_m^\smpl$ and $Q_{p-m}^\smpl$.  In the middle we
must use a new intertwiner of the form
\eqn\KPga{Q_{m+kp}^\smpl=\con{(s_+)^{m+kp}}_{\Om_{m+kp}}\quad {\rm or}
\quad
Q_{(p-m)+kp}^\smpl=\con{(s_+)^{(p-m)+kp}}_{\Om_{(p-m)+kp}}\,,\quad
k\geq1\,,} where $\Om_{m+kp}=\Ga_m\cup\widehat \Ga_p\cup\ldots\cup
\widehat \Ga_p$ and $\Om_{(p-m)+kp}= \Ga_{(m-p)}\cup\widehat
\Ga_p\cup\ldots\cup\widehat \Ga_p$.  The resulting extension of the
Fock space resolution to modules outside the fundamental range can be
summarized by the following theorem.\thm\KPTfext

\proclaim Theorem \KPTfext. Let $1\leq m\leq p-1$, $1\leq m'\leq p'-1$,
and  $j\geq 0$. Then the complex of Fock spaces
\eqn\KPgb{
\matrix{\cdots&
&\mapright{Q_m^\smpl}&
\cF_{-m+(j+2)p,m'}&\mapright{Q_{(p-m)+jp}^\smpl}&\cF_{m-jp,m'}&
\mapright{Q_m^\smpl}&\cF_{-m-jp,m'}&\mapright{Q_{p-m}^\smpl}&\cdots\,,\cr}}
is a resolution of the irreducible module $\cL_{m-jp,m'}$, while
\eqn\KPgc{
\matrix{\cdots&\mapright{Q_m^\smpl}&\cF_{-m+(j+2)p,m'}&
\mapright{Q_{p-m}^\smpl}&\cF_{m+jp,m'}&\mapright{Q_{m+jp}^\smpl}&
\cF_{-m-jp,m'}&\mapright{Q_{p-m}^\smpl}&\cdots\,,\cr}}
is a resolution of $\cL_{m+jp,m'}$.

\proof Contours $\Om_{(p-m)+kp}$ and $\Om_{m+kp}$ are nontrivial cycles
in local homology. This can be verified by a computation similar to the
one in the proof of Lemma \KPTtech. Thus the intertwiners
$Q_{(p-m)+jp}^\smpl$ and $Q_{m+jp}^\smpl$ are precisely those given in
Theorem \KPTinter. In particular they satisfy i) and ii), respectively,
which is precisely what one needs to extend the proof of
Theorem \KPTfeld\ to the present case. \hfill\Box

It is clear that the three other resolutions discussed in Section 4 -- the
dual, and the two others constructed with $Q_\star^\smmi$ instead of
$Q_\star^\smpl$ -- may be extended in the same way to modules outside
the fundamental range.

\newsec{Concluding remarks}

Formally, the new intertwiners in Sections 5, 6 and 7 are proportional to
operators of the form
\eqn\KPgl{{\con{(s_\pm)^k}_\Ga\over[k]_{q_\pm}\,!}\,,}
where $[n]_{q_\pm}$ denotes the usual
$q_\pm$-number. As has been extensively discussed for free field
realizations during the past two
years, the screening currents inside $\con{\cdot}_\Ga$ satisfy the
defining relations of  generators of a quantum group
$\cU_{q_\pm}({\bf n_\pm})$ (see \eg [\BMPrev] and references therein).
It is worth noting that in the discussion of the general class of
resolutions in this letter one is automatically led to the
``rescaled quantum group'' generators of Lusztig [\Lu].

An interesting application of these resolutions is to extend the
computation of the BRST cohomology of minimal models coupled to $2D$
quantum gravity [\LZ] to generalized Dotsenko-Fateev models using
methods discussed  in [\BMPgrav].   This is elaborated in more detail
in [\BMPprep].

\footatend\immediate\closeout\rfile\writestoppt
\baselineskip=14pt{\bigskip\noindent {\bf  References}}%
\bigskip{\frenchspacing%
\parindent=20pt\escapechar=` \input refs.tmp\vfill\eject}\nonfrenchspacing

\vfil\eject\end